\documentclass[twocolumn,aip,reprint,numerical]{revtex4-2}

\usepackage{amsfonts}
\usepackage{graphicx}
\usepackage{epstopdf}
\usepackage{epsfig}
\usepackage{amsmath}
\usepackage[normalem]{ulem}
\usepackage{multirow}
\usepackage{color}
\usepackage{makecell}
\usepackage{array}
\usepackage{booktabs}
\usepackage{mathtools}
\usepackage{bm}
\usepackage{color}
\usepackage{array}
\usepackage{booktabs}
\usepackage{tabularx}

\newcolumntype{Y}{>{\centering\arraybackslash}X}

\newcommand{\Rmnum}[1]{\expandafter\@slowromancap\romannumeral #1@}
\makeatother

\begin{document}

\title{Deep-learning design of graphene metasurfaces for quantum control and Dirac electron holography}

\author{Chen-Di Han}
\affiliation{School of Electrical, Computer and Energy Engineering, Arizona State University, Tempe, Arizona 85287, USA}

\author{Li-Li Ye}
\affiliation{School of Electrical, Computer and Energy Engineering, Arizona State University, Tempe, Arizona 85287, USA}

\author{Zin Lin}
\affiliation{Bradley Department of Electrical and Computer Engineering, Virginia Tech Research Center, 900 N. Glebe Rd, Arlington, VA 22203, USA}

\author{Vassilios Kovanis}
\affiliation{Bradley Department of Electrical and Computer Engineering, Virginia Tech Research Center, 900 N. Glebe Rd, Arlington, VA 22203, USA}
\affiliation{Virginia Tech National Security Institute, Research Center, 900 North Glebe Rd., Arlington, VA 22203, USA} 

\author{Ying-Cheng Lai} \email{Ying-Cheng.Lai@asu.edu}
\affiliation{School of Electrical, Computer and Energy Engineering, Arizona State University, Tempe, Arizona 85287, USA}
\affiliation{Department of Physics, Arizona State University, Tempe, Arizona 85287, USA}

%\date{\today}
\begin{abstract}

Metasurfaces are sub-wavelength patterned layers for controlling waves in physical systems. In optics, metasurfaces are created by materials with different dielectric constants and are capable of unconventional functionalities. We develop a deep-learning framework for Dirac-material metasurface design for controlling electronic waves. The metasurface is a configuration of circular graphene quantum dots, each created by an electric potential. Employing deep convolutional neural networks, we show that the original scattering wave can be reconstructed with fidelity over 95\%, suggesting the feasibility of Dirac electron holography. Additional applications such as plane wave generation, designing broadband and multi-functionality graphene metasurface systems are illustrated.

\end{abstract}

\date{\today}

\maketitle

\section{Introduction} \label{sec:intro}

Metasurfaces are two-dimensional (2D) arrays of subwavelength scatters. A common 
form of metasurfaces is metallic or dielectric structures for modulating or 
controlling electromagnetic waves to achieve desired wavefront, polarization 
distribution, intensity distribution or spectrum~\cite{kildishev2013planar,yu2014flat,chen2016review,lalanne2017metalenses,kamali2018review,shaltout2019spatiotemporal}. 
Optical metasurfaces have wide applications such as planar lens and 
axicons~\cite{aieta2012aberration,ni2013ultra}, vortex 
generators~\cite{genevet2012ultra}, beam deflectors~\cite{shalaev2015high}, and 
holography~\cite{ni2013metasurface,huang2018metasurface}. Compared with 
transform optics that requires continuous changes in the refractive 
index~\cite{pendry2006controlling}, metasurfaces contain distinct elements 
and are experimentally feasible. Metasurface design is important for problems 
such as surface plasmon polarization control, phase and amplitude 
reconstruction~\cite{dolev2012surface,xu2017polarization}, and metasurfaces have 
been exploited for acoustic~\cite{yang2016metasurface,assouar2018acoustic} and 
water surface wave~\cite{sun2016modulation,zou2019broadband} devices as well. 
Metasurfaces can also be extended to matter waves. For example, in electron 
holography, information about the electron wave can be stored and 
reconstructed~\cite{tonomura1987applications,lichte2007electron}, and the
reconstructed wave can provide significantly improved 
resolution~\cite{tonomura1979high,orchowski1995electron}.

An active area of research is to design metasurfaces according to specific 
goals~\cite{elsawy2020numerical}. A previous method was based on impedance 
retrieval that utilizes the local periodicity without 
optimization~\cite{wan2014simultaneous,martini2015metasurface}, 
requiring large metasurfaces that overlap with the target region. A recent trend 
is to exploit machine learning~\cite{so2020deep,wiecha2021deep}, where a rigorous 
solver of the metasurface scattering physics is approximated and replaced by a 
back-propagation type of neural network for high computational efficiency (e.g., 
thousand times faster than traditional optimization 
method)~\cite{peurifoy2018nanophotonic}. Another advantage of machine-learning 
design lies in its tolerance to constraints, in contrast to traditional 
optimization that relies on explicit but often unavailable constraints. For 
example, for designing devices with parameters in a fixed region by the 
interior-point method~\cite{peurifoy2018nanophotonic}, generative 
adversarial neural networks~\cite{liu2018generative} or physics-enhanced machine 
learning~\cite{jiang2019global,chen2022high} can be effective. For discrete
target space, reinforcement learning can be used for optical metasurface
design~\cite{sajedian2019optimisation}. Quite recently, physics-informed neural 
networks for quantum control have been articulated~\cite{NMGC:2024}.

In this paper, we address metasurface design for electronic waves in graphene, 
a widely studied 2D Dirac material~\cite{novoselov2004electric,novoselov2005two,gusynin2005unconventional,rycerz2007valley,han2014graphene,cao2018unconventional,cao2018correlated,neto2009electronic}.
Under the continuum approximation, graphene is effectively a thin conducting 
layer~\cite{gusynin2006unusual,mikhailov2007new,falkovsky2008optical}, and a 
graphene metasurface can be used to manipulate electromagnetic 
waves~\cite{fallahi2012design,miao2015widely,guo2016broadband}, with gate 
potentials generating material layers of different refractive 
indices~\cite{allain2011klein,ozawa2017klein,han2018decay}, thereby offering
more flexibility than optical metasurfaces whose properties are fixed once 
designed. Experimentally, the required gate-potential profiles can be created by 
STM tip~\cite{zhao2015creating,lee2016imaging,gutierrez2016klein,ghahari2017off} 
or doping~\cite{velasco2016nanoscale}. For example, a periodic scattering 
structure leading to a graphene superlattice was realized~\cite{bai2007klein,tiwari2009tunable,burset2011transport,killi2011band}, with experimentally observed 
band structure~\cite{ponomarenko2013cloning}.
%More recently, the Moir\'{e} super lattice was experimentally realized with twisted graphene layers at certain ``magic'' angles and was demonstrated to exhibit superconductivity~\cite{cao2018unconventional,cao2018correlated}. 
A configuration of graphene scatters can generate complex scattering 
phenomena~\cite{sadrara2019dirac}, and there were experiments on multiple 
graphene quantum dots formed by proper electric 
potential~\cite{fu2020relativistic,yang2022creating}.

We focus on designing graphene metasurface to control and manipulate Dirac 
electron scattering to generate any desired wavefront. Consider a point source 
emitting Dirac electronic waves with different energies through scattering from
a graphene metasurface. For convenience, the region of observation, or the target 
region, is a rectangle whose side is approximately twice the wavelength. To be 
concrete, we assume that the metasurface consists of a small number of, e.g., 
six circular scatterers (quantum dots) of a fixed radius. In the language of 
Dirac electron optics, the design goal is to find the value of the dielectric 
constant of each quantum dot to generate any desired electronic waveform in the 
rectangular region of observation. This is an inverse-design problem of finding 
the optimal combination of the gate potentials applied to the quantum dots on 
the metasurface. Because of the need to test a large (infinite in principle) 
number of parameter combinations, optimization based on a rigorous Dirac equation 
solver is computationally infeasible. Moreover, estimating the derivatives of 
the solutions for optimization is challenging due to the need of calculating the 
inverse matrix at each time step. Our solution is to exploit deep convolutional 
neural networks (DCNNs) as a simulator for solving the Dirac 
equation~\cite{ronneberger2015u,he2016deep,HL:2022}. We demonstrate that a 
designed metasurface of as few as six quantum dots can generate rich types of 
scattering wave. A phenomenon is that, given an actual scatterer 
with a complicated geometry, e.g., a star-like scatterer, a DCNN-designed 
metasurface can generate essentially the identical waveform in the target region, 
realizing Dirac electron holography. As will be demonstrated, given an 
optimization goal, it is even possible to generate broadband holography and 
multi-functional devices. Compared with optical metasurfaces whose physical 
parameters are fixed at the creation of the device, graphene based metasurfaces 
have the advantage of flexibility in that the physical parameters can be readily 
modified through the external gate potentials. 

\section{Graphene metasurface and machine-learning design} \label{sec:method}

\subsection{Scattering physics from a graphene metasurface} \label{subsec:graphene}

\begin{figure} [ht!]
\centering
\includegraphics[width=\linewidth]{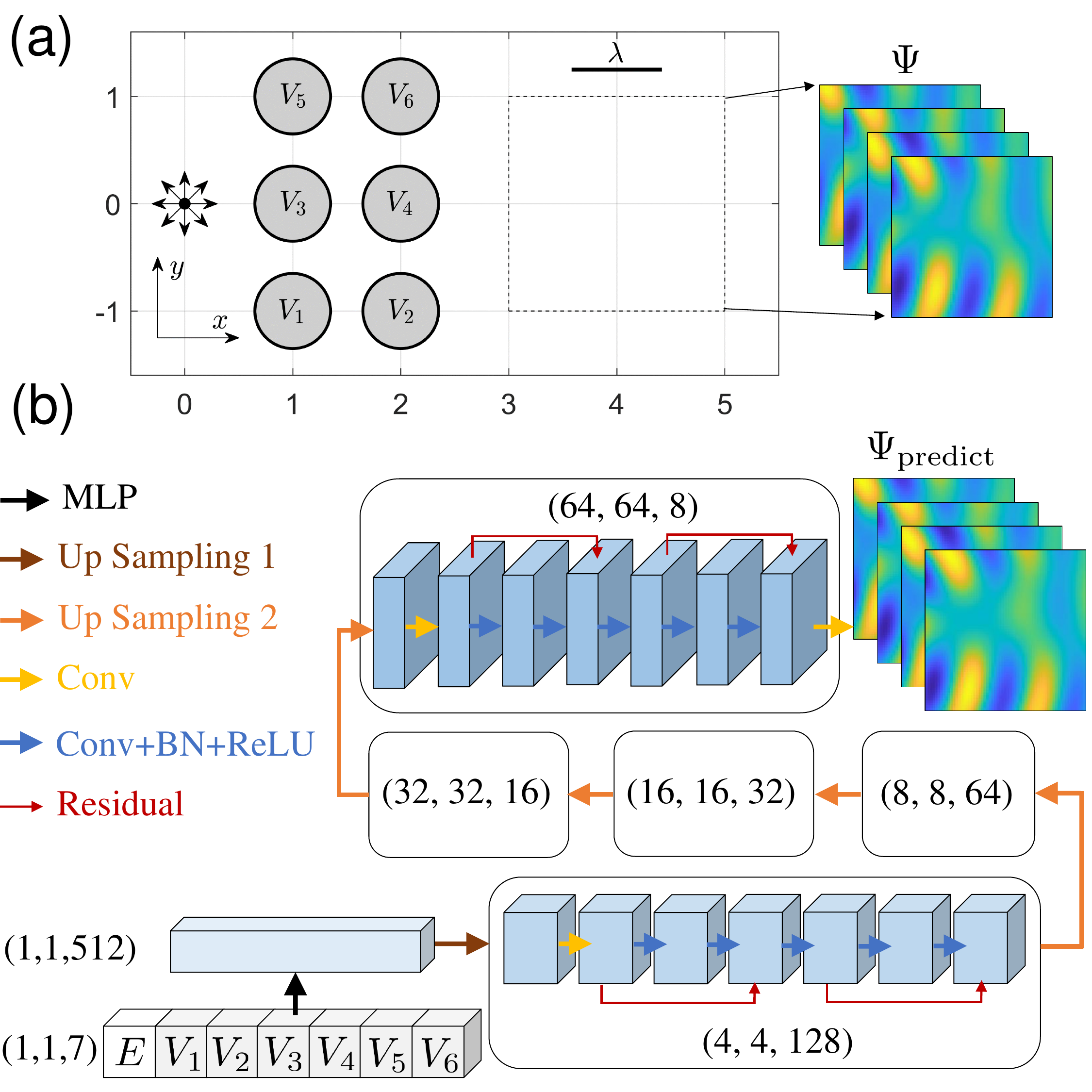}
\caption{Structure of graphene metasurface for quantum scattering and 
machine-learning design. (a) A graphene metasurface consisting of six 
scatterers labeled as $V_1$ to $V_6$, where a cylindrical wave of energy $E$ 
propagates from the left. The square target region on the right is indicated 
as the dashed box. The output spinor wavefunction contains two complex 
components. The four images shown are the real and imaginary parts of the two
components in the order: 
$\Psi=[\text{Re}\psi_1,\text{Im}\psi_1,\text{Re} \psi_2,\text{Im}\psi_2]$. 
(b) A U-net type of convolutional neural network architectures employed 
in our study, where different arrows indicate interlayer transforms. Each 
rounded rectangle represents one convolutional block of certain size. 
Up-samplings connect different blocks. The output images contain four channels 
with the same dimension as that of the spinor wave function.}
\label{fig:schematic}
\end{figure}

We consider a representative graphene metasurface system consisting of six 
circular scatterers, as shown in Figure.~\ref{fig:schematic}(a). In the 
single-electron framework, the Hamiltonian is 
\begin{align} \label{eq:Hamiltonian}
H=v_g\bm{\sigma}\cdot \mathbf{p}+\sum_{i=1}^6 V_i(\mathbf{r})
\end{align}
where $v_g$ is the Fermi velocity, $\bm{\sigma}$ is the vector of Pauli 
matrices, $\mathbf{p}$ is the momentum of the electron, and $V_i(\mathbf{r}$ 
is the potential energy profile that defines the $i$th scatterer, for 
$i=1,\ldots, 6$. For convenience, we use dimensionless units by setting 
$v_g \equiv 1$. The six circles are identical and have the (dimensionless) 
radius $r=0.35$, where the center of each circle is located in $x\in \{1,2\}$ 
and $y\in \{-1,0,1\}$ and each potential-energy $V_i(\mathbf{r}$ is created 
by a proper external gate potential. As a result, the geometric structure and
the physical properties of a graphene metasurface are readily experimentally 
controllable, in contrast to optical metasurfaces defined by the material
properties such as the dielectric constants that cannot be changed once chosen.

For electron holography in three dimensions (3D), energy conservation 
stipulates that the amplitude of a planar incidence wave decays inversely 
with the distance to the origin~\cite{barton1991removing}. For a 2D graphene 
metasurface, the wave function can be written as a two-component spinor 
$\psi=[\psi_1,\psi_2]^T$, where both components $\psi_1$ and $\psi_2$ are
complex. The same energy consideration requires that the wave decays with the 
distance $r$ to the source as $1/\sqrt{r}$. A typical incident spinor wave can 
then be chosen as
\begin{align} \label{eq:incident_wave}
\psi_\text{in}=\frac{1}{\sqrt{2}}\binom{H^{(1)}_0(\mathbf{r})}{iH^{(1)}_1(\mathbf{r})e^{i\theta}}
\end{align}
where $H^{(1)}(\mathbf{r})$ is the Hankel function of the first kind. We choose
the post-scattering target region to be a square, as shown by dashed box 
defined as $x \in [3, 5]$ and $y \in [-1, 1]$ in Fig.~\ref{fig:schematic}(a). 
For machine learning and loss-function computation, real quantities are 
required, so we introduce the following equivalent real spinor wave function: 
$\Psi=[\text{Re}\psi_1,\text{Im}\psi_1,\text{Re}\psi_2,\text{Im}\psi_2]^T$.

To solve the scattering problem, we use the multiple multipole (MMP) method
originated from optics~\cite{LB:1987,Imhof:1996,KA:2002,MEHV:2002,TE:2004}
and adopted to photonic crystal waveguides~\cite{MEH:2002} and surface
plasmons in metallic nanostructures~\cite{MEHV:2002}. The method has recently
been extended to Dirac-Weyl spinor systems under different geometrical and 
mass settings~\cite{XL:2019,XL:2020,WXL:2020,HXL:2020}. For a fixed set of
gate potentials and incident energy, the resulting normalized scattering 
wave in the target region can be calculated using the MMP method, generating
a forward solution.

Our inverse design addresses the problem of identifying an optimal combination
of the electrical potentials applied to the scatterers to achieve a desired
scattering wave, denoted as $\Psi_\text{target}$. To goal is to solve the 
following optimization problem:
\begin{align} \label{eq:inverse_MMP}
\min_{\mathbf{V}} \| \Psi(\mathbf{V})-\Psi_\text{target} \|^2
\end{align}
where $\mathbf{V}$ denotes the set of gate potentials and $\Psi(\mathbf{V})$ 
is a forward solution. Obtaining an optimal solution of $\mathbf{V}$ requiring
repeated use of the MMP method, one use for each potential configuration with
variations determined by a gradient, which is computationally costly. Another
difficulty lies in finding the gradient, which requires matrix inverse 
associated with the MMP method, which can be computationally extremely 
challenging. These difficulties motivated us to exploiting machine learning 
by using convolutional neural networks to approximate $\Psi(\mathbf{V})$.

In optical metasurface design, two different types of problems often arise. 
One type concerns creating a metasurface to transfer energy, such as a planar 
lens~\cite{aieta2012aberration, ni2013ultra} or a beam 
deflector~\cite{shalaev2015high}. For this type of problems, the relevant 
physical quantity is the absolute light intensity or strength. The second 
type is holography~\cite{huang2018metasurface}, which requires reconstructing 
the phase and amplitude of the wave in some target region. For holography, the
relative light strength is relevant since not all the input light energy can
be used to construct a holographic object. In this case, the scattering wave
inside the target region can be normalized to generate data of suitable scale 
for machine-learning design. 

\subsection{Deep convolutional neural networks} \label{subsec:NN}

The basic idea of machine-learning design of graphene metasurface is to use
a DCNN to substitute the Dirac equation solver (MMP). The input to the DCNN is 
the structure of the metasurface characterized by a vector of the gate 
potentials. The output of the DCNN is the spatial distribution of the spinor 
wave function. Since the wave functions spatially adjacent to each other are 
correlated as governed by the Dirac equation, DCNN trained to capture such 
correlations can be used to execute the same function as the Dirac equation.
Such a neural network is equivalent to the inverse version of the neural 
networks typically used in image classification~\cite{krizhevsky2017imagenet}.
In particular, for image classification, a decrease in the spatial size is 
accompanied by a simultaneous increase in the channel size. For our inverse 
problem, increasing the spatial size then requires decreasing the channel 
size (to be explained below). In general, using the residual connection and 
batch normalization can improve the performance of deep neural networks and 
reduce overfitting~\cite{he2016deep,ioffe2015batch}.

\begin{table} [ht!]
\caption{Parameters of convolutional layers}
\begin{tabularx}{\linewidth}{lYYY}
\hline\hline
\specialrule{0em}{1pt}{1pt}
Layer name & Kernel size & Padding & Stride \\
\specialrule{0em}{1pt}{1pt}
\hline
\specialrule{0em}{1pt}{1pt}
Up sampling 1 & $4\times 4$ & $0$ & $1$ \\
\specialrule{0em}{1pt}{1pt}
Up sampling 2 & $4\times 4$ & $0$ & $2$ \\
\specialrule{0em}{1pt}{1pt}
Conv & $3\times 3$ & $1$ & $1$ \\
\specialrule{0em}{1pt}{1pt}
\hline\hline
\end{tabularx}
\label{ta:NN_layer}
\end{table}

Figure~\ref{fig:schematic}(b) shows the DCNN structure used in our study. The 
original input is a vector of seven components, one for energy and six for the 
the six gate potential profiles. Since the neural networks require a 
three-dimensional vector as the input, we add two dummy dimensions so the
input is a three-dimensional matrix: $(1,1,7)$, where the first two dimensions
represent the space and the third value denotes the number of channels. After 
a multilayer Perceptron (MLP), the input matrix maps to a 3D matrix of 
dimension $(1,1,512)$. Up sampling with convolutional kernel size $4\times 4$, 
stride $1$, and zero padding is performed next, where the number of channels 
is reduced in each upsampling with a simultaneous increase in the spacial 
dimension. The next is a convolutional block containing several convolutional 
layers $3\times 3$, stride $1$ and $1$ padding. It also contains residual 
connections and batch normalization (BN)~\cite{ioffe2015batch}. After the first
block, another up sampling is performed with convolutional kernel size 
$4\times 4$, stride $2$, and $0$ padding, converning the 3D matrix into 
one of dimension $(8,8,64)$. A similar structure is repeated three times. 
Finally, when the convolutional block of spacial dimension $64$ is done, we 
use a 2D convolutional layer and set the output dimension to be $(64,64,4)$. 
Table~\ref{ta:NN_layer} summarizes the three different convolutional layers 
used in our study. It is worth noting that, even with the same convolutional 
layer, the number of channels can be different for different blocks.

It is worth comparing our DCNN architecture with two typical neural networks
for inverse design. First, compared with the traditional generative adversarial 
neural network (GAN) that was recently adopted for inverse 
design~\cite{liu2018generative}, our DCNN architecture contains a residual 
connection specifically for complicated training data, e.g., output wave
function patterns from different energy values. Second, we note that the U-Net 
originally introduced for inverse design of medical images does not contain 
shortcut connections~\cite{ronneberger2015u}.

The training data are generated, as follows. We first set (quite arbitrarily) 
the energy range to be $E\in [5,10]$ and choose the gate potential such that 
$V_i \in [-5,20]$ (these input data to the neural networks are normalized 
to the unit interval). As shown in Fig.~\ref{fig:schematic}(a), the size of the 
square target region to observe the scattering patterns of incident cylindrical 
waves is approximately two or three times of the incident wavelength. We then
use the MMP method to calculate the scattering wave functions (see 
Appendix~\ref{sec:Appendix_A} for details). The target region is discretized 
into a $64\times 64$ grid and the wave function at each grid point is obtained. 
Finally, we normalize the wave function in the target region to generate the 
training data 
$\Psi=[\text{Re}\psi_1,\text{Im}\psi_1,\text{Re }\psi_2,\text{Im}\psi_2]$.
Altogether, we generate $80,000$ training data sets (scattering patterns) from 
randomly generated energy and potential values in their respective ranges. 
For comparison, with brute-force sampling, since the input data is a 
seven-dimensional vector, $80,000$ training data sets are equivalent to taking
only five points in each dimension.  

\begin{table} [ht!]
\caption{Values of the rraining parameters for DCNN}
\begin{tabularx}{\linewidth}{YY}
\hline\hline
\specialrule{0em}{1pt}{1pt}
Description & Values \\
\specialrule{0em}{1pt}{1pt}
\hline
\specialrule{0em}{1pt}{1pt}
Batch size &  $128$ \\
\specialrule{0em}{1pt}{1pt}
Learning rate &  $0.005$ \\
\specialrule{0em}{1pt}{1pt}
Optimizer &  Adam \\
\specialrule{0em}{1pt}{1pt}
Number of epochs &  $1000$ \\
\specialrule{0em}{1pt}{1pt}
Number of training data &  80,000 \\
\specialrule{0em}{1pt}{1pt}
Number of validating data &  10,000 \\
\specialrule{0em}{1pt}{1pt}
Number of testing data &  10,000 \\
\specialrule{0em}{1pt}{1pt}
\hline\hline
\end{tabularx}
\label{ta:NN_hyperparameter}
\end{table}

\begin{figure}
\centering
\includegraphics[width=\linewidth]{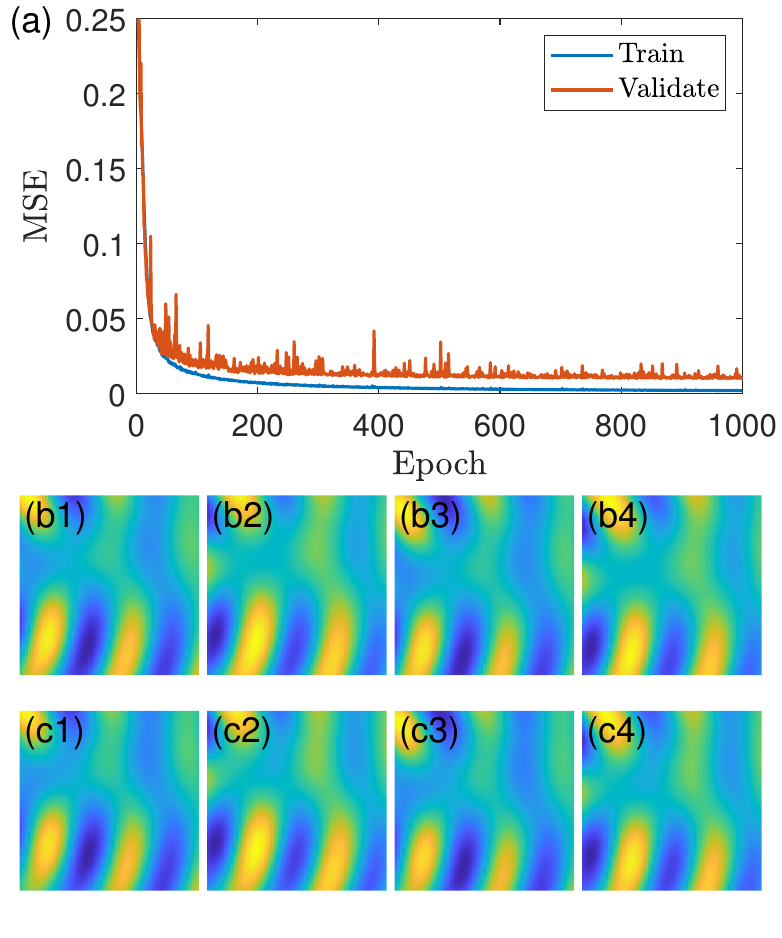}
\caption{Performance of DCNN training. (a) Training and validation errors versus
epoch number. The errors converge to small values after about $1000$ epochs. 
(b1-b4) The true wavefunction calculated by the MMP method, where the four 
images from left to right correspond to the real spinor components Re $\psi_1$, 
Im $\psi_1$, Re $\psi_2$, and Im $\psi_2$, respectively. (c1-c4) The 
corresponding DCNN predicted spinor components, where the MSE is less than
0.01, similar to the averaging testing error.}
\label{fig:loss}
\end{figure}

Table~\ref{ta:NN_hyperparameter} lists the values of the training parameters. 
The loss function is the mean square error (MSE):
\begin{align} \label{eq:Loss}
\mathcal{L}_\text{Train}=\| \Psi_\text{real}-\Psi_\text{predict}\|^2
\end{align}
The neural network is built using PyTorch~\cite{paszke2019pytorch} and GPU is
employed for fast training. Figure~\ref{fig:loss}(a) shows the training and 
validating errors versus the epoch number. The fluctuations are mainly due to 
batch normalization used in the neural network, which can be reduced by using 
a larger batch size. However, a large batch slows down the convergence, so there
is a computational tradeoff. Figures~\ref{fig:loss}(b1-b4) and 
\ref{fig:loss}(c1-c4) show the true (calculated from the MMP method) and DCNN 
predicted scattering wave function from one example in the testing dataset 
whose MSE error is about the average testing error, where the four panels in 
each row correspond to the four real spinor components Re $\psi_1$, Im $\psi_1$,
Re $\psi_2$ and Im $\psi_2$, respectively. Visually, there is little difference 
between the DCNN predicted scattering patterns and the ground truth, indicating 
that effective training has been achieved. Table~\ref{ta:NN_train_result} lists
the MSEs for training, evaluating and testing from the dataset with the smallest
validation error. It can be seen that the training MSE is smaller than the 
validation and testing MSEs, indicating a certain degree of overfitting. Using 
more training data can help mitigating this problem.  

\begin{table} [ht!]
\caption{MSEs from the trained DCNN}
\begin{tabularx}{\linewidth}{YY}
\hline\hline
\specialrule{0em}{1pt}{1pt}
Description & Values \\
\specialrule{0em}{1pt}{1pt}
\hline
\specialrule{0em}{1pt}{1pt}
Training MSE &  $0.0020$ \\
\specialrule{0em}{1pt}{1pt}
Validating MSE &  $0.0101$ \\
\specialrule{0em}{1pt}{1pt}
Testing MSE &  $0.0102$ \\
\specialrule{0em}{1pt}{1pt}
\hline\hline
\end{tabularx}
\label{ta:NN_train_result}
\end{table}

\section{Dirac electron holography}

\begin{figure}
\centering
\includegraphics[width=\linewidth]{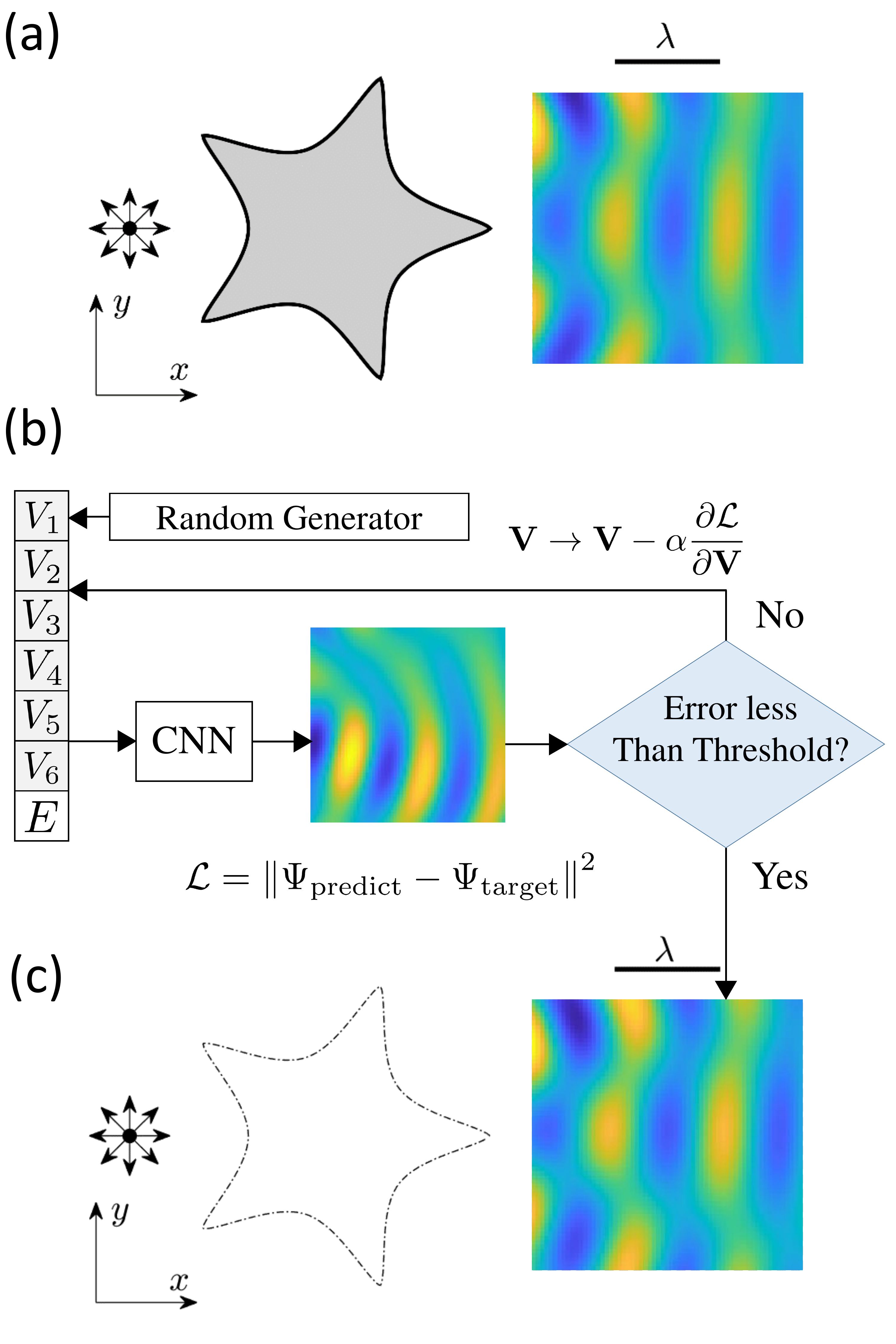}
\caption{Inverse design for Dirac electron holography. (a) A real scatter of a 
star shape with potential $V=20$ (details in Appendix~\ref{sec:Appendix_B}). 
The scattering wave function in the rectangle region is calculated directly 
from the star-shaped potential by using the MMP method. Shown on the right is 
a real component of the scattering wave Re $\psi_1$. (b) Illustration of the 
inverse design process. The target wave and the energy are the inputs to the 
DCNN. Initially, the potential configuration $\mathbf{V}$ is generated randomly.
The DCNN generates the scattering wave and compare it with the target wave in 
(a), and the error is used to modify the potential configuration $\mathbf{V}$. 
The process is iterated until the loss function is below a pre-selected error 
threshold, leading to the optimal potential configuration. (c) Scattering wave 
pattern calculated from the optimal potential configuration through, again, the 
MMP method, which matches the desired wave pattern in (a), thereby creating a 
holographic image of the star-shaped scatterer and realizing Dirac electron 
holography.}
\label{fig:design}
\end{figure}

A well-trained DCNN can substitute the real Dirac equation solver and provide 
an accurate solution in a computationally efficient way. In 
particular, with the available open source packages such as 
PyTorch~\cite{paszke2019pytorch}, computing the loss and the gradient with 
respect to the input on GPU servers can be done extremely efficiently.
For inverse design, the loss function is
\begin{align} \label{eq:inverse_NN}
\mathcal{L}_\text{design}= \| \Psi_\text{predict}(\mathbf{V})-\Psi_\text{target}\|^2
\end{align}
where $\Psi_\text{predict}$ is the output from the neural network and 
$\Psi_\text{target}$ is the desired target wave. Figure~\ref{fig:design} 
presents an example of generating a desired wave pattern, optimization, and
creating Dirac electron holography. In particular, Fig.~\ref{fig:design}(a) 
shows a real scatterer of a star shape generated by the gate potential $V=20$.
For the cylindrical incident wave of energy $E=8$, the scattering-wave pattern 
in the target region, calculated directly from this star scatterer using the 
MMP method (details in Appendix~\ref{sec:Appendix_B}), is also shown. 
Figure~\ref{fig:design}(b) shows the optimization algorithm to minimize the 
loss in Eq.~\eqref{eq:inverse_NN}, which starts from a randomly generated 
potential. We input the initial potential to the neural network and compute 
the loss and the gradient to variations in the metasurface parameter vector 
$\mathbf{V}$, and update $\mathbf{V}$ by the optimization algorithm 
(source package scipy.optimize.minimize~\cite{2020SciPy-NMeth} in Python) 
at certain learning rate $\alpha$ that depends on the specific optimization 
method. Note that, since the electron energy as an input to the neural network 
is given for the design process, we use the potentials to evaluate the gradient.
In fact, with the known target wave pattern, the energy can be obtained from
the Dirac equation. The optimization result depends on the initial condition,
so we use a small ensemble of (ten) initial conditons and choose the potential
configuration $\mathbf{V}$ that yields the smallest value of 
$\mathcal{L}_\text{design}$. This combination of the potential values also
gives the smallest loss in Eq.~\eqref{eq:inverse_MMP} for a properly trained
DCNN. To demonstrate that the so-designed metasurface with the optimal potential
configuration can generate the desired target wave pattern, we again employ
the MMP method to calculate the scattering wave in the target region, but 
this time from the potential configuration. The resulting scattering wave 
pattern in shown in Fig.~\ref{fig:design}(c), which agrees well with that 
calculated from the star scatterer itself in Fig.~\ref{fig:design}(a). The
DCNN generated metasurface can thus faithfully generate the desired scattering
wave pattern from a geometrically complicated scatterer such as a star. 
Conversely, the metasurface generated wave pattern corresponds to the 
specific star scatterer, as shown by the dotted-dashed shape in 
Fig.~\ref{fig:design}(c), realizing Dirac electron holography!

To characterize the design accuracy, we define the following fidelity measure:
\begin{align} \label{eq:fidelity}
F=|\psi_\text{design}\psi_\text{target}^*|,
\end{align}
where $\psi_\text{design}$ is the normalized scattering wave generated by the
metasurface and $\psi_\text{target}$ is the normalized wave from a target 
scatterer as governed by the Dirac equation, for the same energy $E$. In the
ideal case where the metasurface generates a wave that matches perfectly the 
target wave, we have $F=1$ (due to normalization). 

\begin{figure}
\centering
\includegraphics[width=\linewidth]{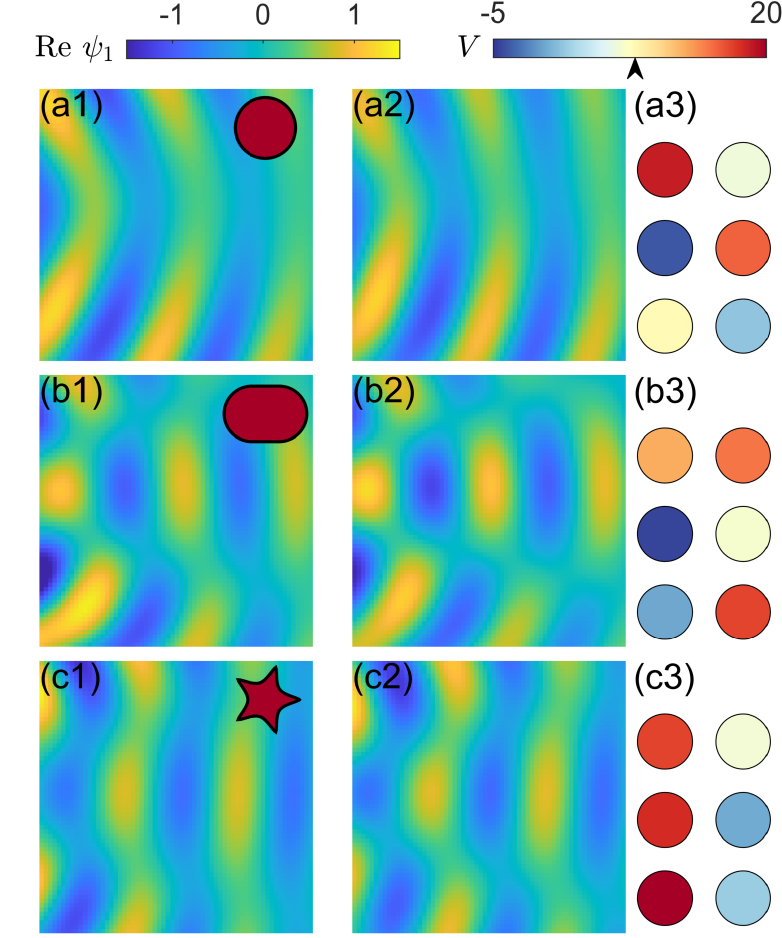}
\caption{Examples of scattering-wave matching and Dirac electron holography 
for three different objects. (a1) Scattering wave from a circular cavity of 
potential $V=20$ with cylindrical incident wave of energy $E=8$, (a2) scattering
wave from the DCNN designed metasurface, and (a3) color-coded potential 
configuration of the designed metasurface. (b1-b3) Same legends as in (a1-a3) 
but for a stadium-shaped holographic object. (c1-c3) Same Same legends as in 
(a1-a3) but for a star-shaped holographic object.}
\label{fig:single_target}
\end{figure}

The general principle of holography is scattering wave matching, as can be seen
by comparing the desired wave generated by a star-shaped scatterer in 
Fig.~\ref{fig:design}(a) and that generated by the designed metasurface in 
Fig.~\ref{fig:design}(c). Two more examples are shown in the upper and middle
rows in Fig.~\ref{fig:single_target}, where the holographic objects have the
shape of a circle and a stadium, respectively (the geometric parameters 
specified in Appendix~\ref{sec:Appendix_B}). The energy values for the two cases 
are $E=8$. For comparison, we also include the case of the star-shaped 
holographic object (shown in Fig.~\ref{fig:design}) in the bottom row of 
Fig.~\ref{fig:single_target}. For the three cases, the desired wave pattens  
are shown in the left column, and those generated by the DCNN-designed 
metasurface are shown in the middle column, with the potential configuration 
of the metasurface in the right column. It can be seen that for the three 
holographic objects, an excellent scattering-wave matching has been achieved.
The resulting fidelity values for three different energy values are listed in
Tab.~\ref{ta:single_target}. In all cases, the fidelity values are larger than
$95\%$. Note that for a larger energy, the fidelity value decreases slightly
due to the target region's containing more wavelengths. 

\begin{table} [ht!]
\caption{Fidelity values for a circular, stadium, and star holographic object 
for different incident energies}
\begin{tabularx}{\linewidth}{YYYY}
\hline\hline
\specialrule{0em}{1pt}{1pt}
Geometric shape & $E=7$ & $E=8$ & $E=9$ \\
\specialrule{0em}{1pt}{1pt}
\hline
\specialrule{0em}{1pt}{1pt}
Circle &  $0.992$ & $0.965$ & $0.963$\\
\specialrule{0em}{1pt}{1pt}
Stadium &  $0.967$ & $ 0.966$ & $0.961$ \\
\specialrule{0em}{1pt}{1pt}
Star &  $0.988$ & $ 0.974$ &  $0.959$ \\
\specialrule{0em}{1pt}{1pt}
\hline\hline
\end{tabularx}
\label{ta:single_target}
\end{table}

There are some constraints for the loss function Eq.~\eqref{eq:inverse_NN},
which leads to a metasurface with specific phase matching. A constant phase 
shift $\phi$ will not only modify the spinors to $\psi \exp(i\phi)$, but also
change the value of the loss function. While phase is important for 
interference-related problems in quantum 
system~\cite{rutter2007scattering, young2009quantum}, it is not crucial for 
physical quantities such as the local electron density, current and pseudospin 
polarization. For those phase-independent quantities, the loss function 
Eq.~\eqref{eq:inverse_NN} may not yield the best structure. The second 
constraint is that the absolute information for the amplitude is missing 
in Eq.~\eqref{eq:inverse_NN} due to normalization. 

It is worth noting that, in optical holography, the energy $E$ corresponds to 
the input frequency and the potential configuration $\mathbf{V}$ represents the 
dielectric property of the metasurface. If the output is a scattering cross 
section or transmission, MLPs are commonly used with the output being a vector 
covering all the frequencies~\cite{peurifoy2018nanophotonic}. For graphene 
metasurface, the aim of our inverse design is to achieve wave function matching,
so the DCNN output is a $2\times 2$ matrix representing the scattering wave 
function. Another difference from optical metasurface is that the electron 
energy value $E$ is also input to the neural network, so the output is the 
wave function but at the specific energy value, enabling faster inverse design.

\section{Other applications of machine-learning designed graphene metasurfaces} \label{sec:applications}

We address three additional applications of graphene metasurfaces. 

\subsection{Plane wave generation} \label{subsec:plane_wave_gen}

An important application of metasurfaces in optical system is to transform a 
cylindrical wave into another type of 
wave~\cite{wan2014simultaneous,martini2015metasurface}. To demonstrate that
this is also possible with our machine-learning based Dirac electron holography,
we begin with a spinor plane wave - a simple solution of the Dirac equation:	
\begin{align} \label{eq:spinor_plane_wave}
\psi_\text{plane}=\binom{1}{\tau}\exp(ikx)
\end{align}
where $\tau=\text{sign}(E)$ and $k=|E|/v_g$. We normalize the wave and set it 
as the desired wave pattern in the target region, and aim to transform an
incident cylindrical wave into this plane wave. Note that, even without 
metasurface scattering, the cylindrical and plane waves share a certain degree 
of similarity, especially in the small energy regime, as indicated in the 
middle column of Tab.~\ref{ta:plane_wave}, where the values of ``natural''
(i.e., without any scatterer) fidelity between the two types of waves for 
several energies are listed. It can be seen that the fidelity value is high
for small energy, and decreases as the energy increases. For a metasurface
to be meaningful, for any energy the achieved fidelity value should be larger 
than the ``natural'' value.

\begin{figure} [ht!]
\centering
\includegraphics[width=\linewidth]{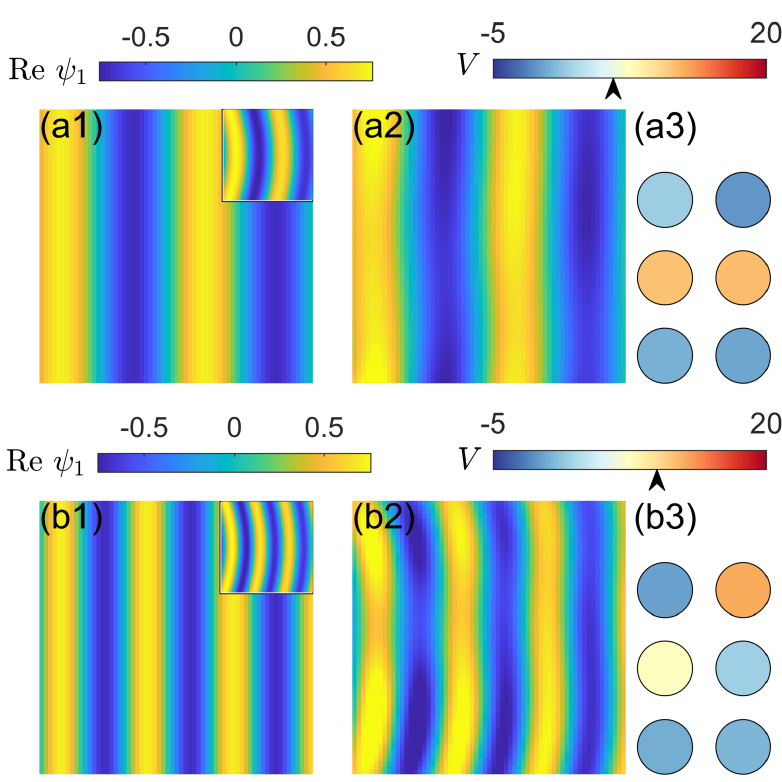}
\caption{Illustration of a graphene metasurface plane-wave generator. (a1) A 
desired plane wave in the target region for $E=6$, where the color represents 
the strength of $\rm{Re}\{\psi_1\}$ and the inset in the top right corner 
displays the incident cylindrical wave at the same energy. (a2) The plane wave 
generated by the DCNN designed metasurface for $E=6$ with the high fidelity 
value $0.996$. (a3) The resulting metasurface of six graphene quantum dots 
with the color representing the potential strength. The black arrow indicates 
the energy value. (b1-b3) Another example: generating a desired plane wave for 
$E=10$. Legends are the same as in (a1-a3). The fidelity value is $0.966$, 
which decreases slightly from the value in (a2), due to the higher energy.} 
\label{fig:plane_wave}
\end{figure}

\begin{table} [ht!]
\caption{Fidelity values for metasurface based graphene plane-wave generator}
\begin{tabularx}{\linewidth}{YcY}
\hline\hline
\specialrule{0em}{1pt}{1pt}
Energy & Fidelity (natural) & Fidelity (designed) \\
\specialrule{0em}{1pt}{1pt}
\hline
\specialrule{0em}{1pt}{1pt}
$5$ & $0.972$ &  $0.988$ \\
\specialrule{0em}{1pt}{1pt}
$6$ & $0.964$ &  $0.996$ \\
\specialrule{0em}{1pt}{1pt}
$7$ & $0.954$ &  $0.989$ \\
\specialrule{0em}{1pt}{1pt}
$8$ & $0.942$ &  $0.973$ \\
\specialrule{0em}{1pt}{1pt}
$9$ & $0.930$ & $ 0.963$ \\
\specialrule{0em}{1pt}{1pt}
$10$ & $0.915$ & $0.966$ \\
\specialrule{0em}{1pt}{1pt}
\hline\hline
\end{tabularx}
\label{ta:plane_wave}
\end{table}

Figure~\ref{fig:plane_wave}(a1) shows the spatial pattern of $\rm{Re}\{\psi_1\}$ 
taken from the plane wave for $E=6$, where the upper-right inset is the 
incident wave. Figure~\ref{fig:plane_wave}(a2) shows the scattering wave from 
our DCNN-generated metasurface whose structure is displayed in 
Fig.~\ref{fig:plane_wave}(a3), where the color for each quantum dot indicates
the potential value (red for high positive and blue for low negative value). 
The black arrow pointing at the upper color bar specifies the value of the
incident energy. The fidelity achieved in this vase is $0.996$. The striking
similarity between the wave patterns in Figs.~\ref{fig:plane_wave}(a1) and 
\ref{fig:plane_wave}(a2) is indicative of the success of the DCNN metasurface 
design. Another example is shown in Figs.~\ref{fig:plane_wave}(b1-b3), for
a higher energy value: $E=10$. The scattering wave pattern from the designed 
metasurface exhibits oscillations, due to the high energy. The third coloum of
Tab.~\ref{ta:plane_wave} lists the fidelity values from the metasurface 
generated plane waves for a number of energy values. As the energy increases,
there is a slow decrease in the fidelity value. This can be explained by noting
that, since the incident cylindrical wave is from the left, the scattering wave 
amplitude can take larger values on the left than on the right side of the 
metasurface scatterer, but an ideal plane wave has the same amplitude at any 
point in the propagation direction [the difference can be seen from 
Figs.~\ref{fig:plane_wave}(b1) and \ref{fig:plane_wave}(b2)]. In addition,
as the energy increases, the wavelength decreases so that the same target
region of observation contains more wavelengths, making it more difficult for
the designed metasurface to generate a plane wave in this region.

\subsection{Designing broadband graphene metasurface systems}

For a metasurface system designed to modulate or control waves, whether optical
or Dirac electron waves, bandwidth is an important characterizing quantity. In 
particular, will the system function as desired in a broad frequency (energy)
range or will it work only for specific frequencies (energies)? The bandwidth
issue is also crucial in other situations such as designing a wave system for 
cloaking or supperscattering in certain frequency (energy) 
range~\cite{monticone2013cloaked,fleury2015invisibility,HL:2022}. As illustrated 
in Fig.~\ref{fig:design}, the geometric shape of our DCNN-designed metasurface 
differs drastically from the actual scatterer. Since the potential configuration
defining a specific metasurface was generated through scattering data at certain
energies, acceptable wave matching (holography) cannot be anticipated to arise
for all energies. Would it be possible to design a metasurface device with the
desired functionality in a limited energy range? To address this question, we 
define the following quantity to characterize the bandwidth:
\begin{align} \label{eq:bandwidth}
\Delta \equiv (\lambda_\text{max}-\lambda_\text{min})/\lambda_\text{min}, 
\end{align}
where $\lambda_\text{min}$ and $\lambda_\text{max}$ are the minimum and maximum 
wavelength and the performance of the metasurface system is acceptable for any
wavelength $\lambda \in [\lambda_\text{min}, \lambda_\text{max}]$. A similar
quantity was introduced, e.g., in a previous work on designing a strong 
scattering system~\cite{peurifoy2018nanophotonic}, where a suitable set of loss 
functions leading to strong scattering was effective even with about a $25\%$
shift in the wavelength. In another work~\cite{devlin2016broadband}, the 
geometrical phase was exploited to tolerate a $100\%$ wavelength shift in the
inverse design. 

\begin{figure} [ht!]
\centering
\includegraphics[width=\linewidth]{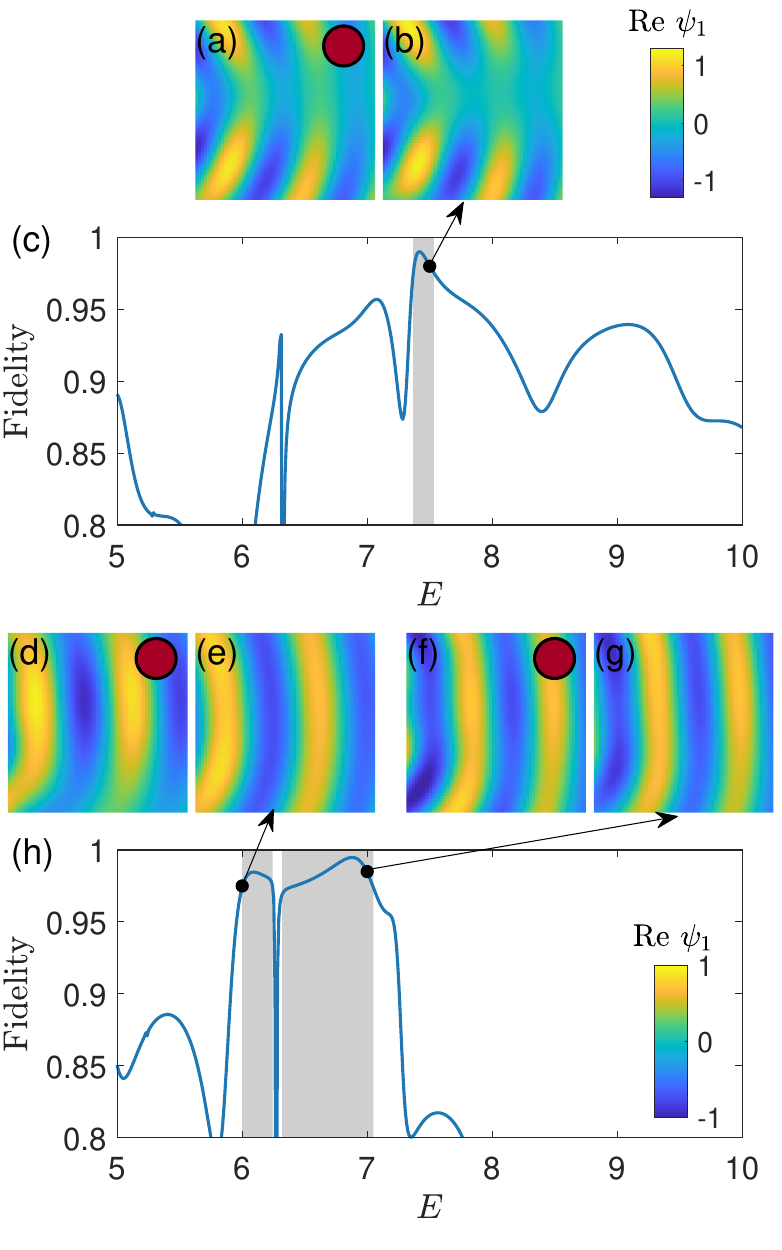}
\caption{Wideband metasurface design illustrated using a circular scatterer.
The circular scatterer is generated by gate potential $V=20$. (a) The scattering
wave pattern $\rm{Re}\{ \psi_1\}$ as a solution of the Dirac equation in the 
target observational region for $E=7.5$. (b) The wave pattern from the metasurface
designed for the same energy. (c) The fidelity versus the energy, where the same 
metasurface (designed for $E=7.5$) is applied to all energy values. The black dot 
indicates the energy $E=7.5$ and the vertical gray strip indicates the energy 
interval in which the fidelity value is larger than $0.975$, which gives the 
bandwidth $\Delta \approx 3\%$ in terms of wavelength. (d) The actual wave 
pattern for $E=6$. (e) The wave pattern generated by the metasurface designed for 
$E=6$. (f) The actual wave pattern for $E=7$. (g) The wave pattern generated by 
the metasurface designed for $E=7$. (h) Fidelity versus the incident energy, 
where the metasurface is designed using two energy values: $E=6$ and $E=7$ 
(indicated by the two black dots). For most energies in between them, the 
fidelity value is high. The corresponding bandwidth has been increased to 
$\Delta \approx 17\%$ of the wavelength.}
\label{fig:circle_multi}
\end{figure}

Our goal is to assess whether Dirac electron holography can be designed to 
function in a relatively broad energy range. To gain insights, we consider a 
circular graphene scatterer. Figure~\ref{fig:circle_multi}(a) shows the 
scattering wave pattern as a solution of the Dirac equation in the observational 
region for $E = 7.5$, where the scatterer is generated by the gate potential 
$V=20$. Figure~\ref{fig:circle_multi}(b) shows the wave pattern from the 
corresponding metasurface at the same energy, which matches the actual pattern 
to a large extent. The question is, if the energy is changed, can a reasonable
match between the two scattering wave patterns hold? To answer this question,
we calculate the fidelity characterizing the wave-pattern matching for a wide
energy range, as shown in Fig.~\ref{fig:circle_multi}(c), where the metasurface
is designed for $E=7.5$ (indicated by the filled black circle) for which the 
fidelity value is $0.980$. The vertical gray strip indicates the energy 
interval in which the fidelity is higher than $0.975$. As the energy deviates
from this interval, the fidelity value decreases rapidly, so the gray interval
represents the bandwidth of the specific design, which is quite narrow 
relatively as it indicates that the design can tolerate only $\Delta = 3\%$ of 
the wavelength change.

We use the method of interval-training~\cite{peurifoy2018nanophotonic} to increase
the bandwidth of the designed metasurface. The idea is to use the wave patterns
in an energy interval $[E_1,E_2]$ for inverse design. The corresponding loss 
function is 
\begin{align} \label{eq:Loss_Interval}
\mathcal{L}_\text{design}=\int_{E_1}^{E_2} \| \Psi_\text{predict}(E, \mathbf{V})-\Psi_\text{target}(E)\|^2 dE.
\end{align}
If the energy interval $[E_1,E_2]$ is relatively narrow, the loss function can
be approximated by
\begin{align} \label{eq:Loss_E1_E2}
\begin{split}
\mathcal{L}_\text{design}=& \| \Psi_\text{predict}(E_1, \mathbf{V})-\Psi_\text{target}(E_1)\|^2\\
&+\| \Psi_\text{predict}(E_2, \mathbf{V})-\Psi_\text{target}(E_2)\|^2.
\end{split}
\end{align}
To test the loss function in Eq.~\eqref{eq:Loss_E1_E2}, we set $E_1=6$ and 
$E_2=7$, and then minimize the loss to find the potential configuration 
$\mathbf{V}$. The true and metasurface-generated scattering wave patterns for
the two energies are shown in Figs.~\ref{fig:circle_multi}(d-g), and  
Fig.~\ref{fig:circle_multi}(h) shows the fidelity versus the energy for
$E \in [5,10]$, where the two black dots indicate the two energy values 
required by the loss function Eq.~\eqref{eq:Loss_E1_E2}. For most energy values 
in the vertical gray shaded regions (except for a sharp drop about $E=6.3$, the 
fidelity value is larger than $0.975$, giving the wavelength bandwidth 
$\Delta \approx 17\%$ - a significant improvement compared with that from a 
single energy loss function, as shown in Fig.~\ref{fig:circle_multi}(c). 

\subsection{Designing multi-functionality graphene metasurface}

\begin{figure} [ht!]
\centering
\includegraphics[width=\linewidth]{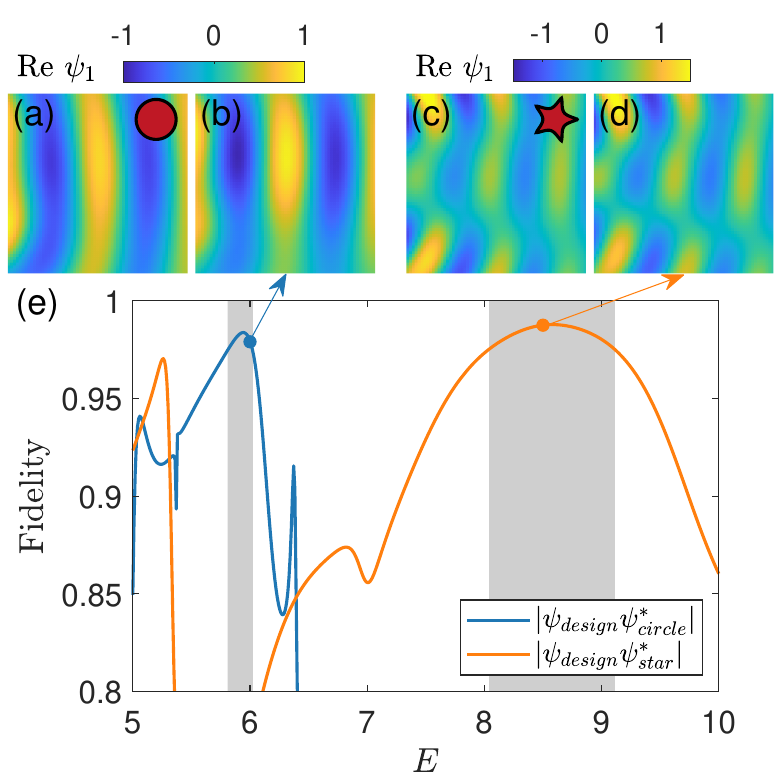}
\caption{An example of designing graphene metasurface with multiple 
functionalities. (a) Desired scattering wave from a circular scatterer for 
$V=19$ for $E=6$. (b) The corresponding wave pattern generated by the 
metasurface for $E=6$. (c) Desired wave pattern from a star-shaped scatterer 
for $V=19$ at $E=8.5$. (d) The corresponding wave pattern generated by the 
metasurface for $E=8.5$. (e) Two types of fidelity versus the energy: 
blue (orange) trace - fidelity between the desired wave from the circular 
(star-shaped) quantum dot and the corresponding metasurface-generated wave. 
The vertical gray shading strips indicate the energy interval with fidelity 
higher than $0.975$. The designed metasurface is capable of double-functional 
holography in two different energy intervals.}
\label{fig:circle_star}
\end{figure}

In optics, designing metasurfaces that can perform multiple functions was 
investigated~\cite{verslegers2012electromagnetically}. For example, complicated 
metasurface can generate different holographic patterns at different 
frequencies~\cite{zhu2021building}. Factors that can change the device properties
include phase transition~\cite{huang2018switching,lepeshov2019nonscattering}
and biases~\cite{farhat20133d,li2015atomically}. Here we address the problem
of designing multi-functionality graphene metasurfaces through some proper gate
potential configuration $\mathbf{V}$. To illustrate our approach, we consider 
two energy values $E_1$ and $E_2$, where the desired scattering wave patterns 
for the two energy values are different. For example, for $E_1$, the desired
wave pattern should match that from a circular quantum dot, while for $E_2$, 
the metasurface generated wave pattern should approxiate that from a star-shaped
quantum dot, as shown in Figs.~\ref{fig:circle_star}(a-d). The loss function is 
\begin{align}
\begin{split}
\mathcal{L}_\text{design}=& \| \Psi_\text{predict}(E_1, \mathbf{V})-\Psi_\text{target1}\|^2\\
&+\| \Psi_\text{predict}(E_2, \mathbf{V})-\Psi_\text{target2}\|^2.
\end{split}
\end{align}
As a concrete example, we set $E_1=6$ and $E_2=8.5$ and perform optimization 
of the loss function to obtain the optimal potential configuration $\mathbf{V}$. 
Figure~\ref{fig:circle_star}(e) shows the fidelity versus the energy $E$, where
the blue (orange) trace is the fidelity with respect to the scattering wave from 
the circular (star-shaped) quantum dot, with the respective shaded strips in which
the fidelity value is larger than $0.975$. The result suggests that two different
functionalities can be realized by the designed graphene metasurface in two 
different energy intervals, respectively. Empirically, the two energy values 
should not be too close to each other. Also, assigning a higher energy value 
for more complicated scattering pattern (e.g., star-shaped dot) can be beneficial
for the optimization process. An intuitive reason is that scattering patterns are 
generally more complicated at high energies, so using a complicated scatterer at
a high energy value can provide more complexity to the neural network to enhance
its computational capability and to generate a broad range of solutions of the 
Dirac equation.

\section{Discussion} \label{sec:discussion}

Optical metasurfaces were invented to manipulate the wavefront and create  
holography. Naturally, the concept can be extended to other wave systems, such 
as the low-energy excitations in graphene governed by the Dirac equation. 
Existing studies of graphene metasurface focused mostly on its optical properties,
i.e., its use as a dielectric medium to modulate electromagnetic waves. Whether
graphene metasurfaces can be used to control electronic waves and to create 
Dirac electron holography remained to be an open question. In the present work,
we addressed this question by developing a deep learning based, inverse-design
framework to generate ``electronic'' graphene metasurfaces. The prototypical
type of metasurface in our study consists of a small number of quantum dots on
a graphene sheet, which can be realized through external electric gate voltages.
Especially, the voltages applied to the quantum dots are different and constitute
a set of parameters that can be optimized through machine learning. We 
demonstrated, using a graphene metasurface of six quantum dots, that various 
desired electronic wave patterns can be generated through quantum scattering and 
Dirac electron holography can be realized. 

Our machine-learning design is to train a DCNN trained to generate a relation
between a set of device parameters (e.g., the gate voltages) and a desired 
waveform that can be a plane wave or the scattering wave from a scatterer 
with a particular geometric shape. Training is done in an ``offline'' fashion 
to generate the required parameters for the metasurface. In the case of a single 
target wave, the wave generated by the metasurface can match the desired wave 
with fidelity higher than $95\%$. Such a high-fidelity wave matching is 
essentially what is required for producing Dirac electron holograph. For 
metasurfaces generated from training data from a single energy value, waveform 
matching can be achived in a small interval about this energy value. To increase 
the energy (frequency) band, we articulated a loss function that involves the 
desired and deep-learning predicted wave functions at multiple energy values. 
Dual functionalities were also demonstrated where a metasurface system can realize 
Dirac electron holography at two distinct energy values. 

One requirement in our framework is that the designed metasurface functions
for the same energy values used to generate the waveforms for training. In 
machine learning, this is referred to as the problem of ``overlapping of the 
design space with the training space.'' This is in fact a common difficulty 
for inverse design of wave scattering systems. Another challenge is that, during
the training process, the DCNN are updated iteratively, which requires a large
computational load. Use of generative adversarial neural 
networks~\cite{liu2018generative} can reduce the computations, as it directly
finds a relation between the desired performance and the device parameters. 
A difficulty is that the devices parameters so produced can often be unphysical,
requiring some sophisticated filtering process to obtain physical reasonable
parameter values.

The capacity achieved in this paper for controlling Dirac electron waves through 
graphene metasurfaces is relatively small compared to what metasurfaces can do to 
optical waves, for two reasons. First, scattering occurs in 2D so the wave 
amplitude on the side of the target observational region closer to the source 
is larger than that on the opposite side. Second, electron scattering in Dirac 
systems is generally weak due to Klein tunneling. These two factors limite the 
type of waveforms that can be generated by a graphene metasurface. Nonetheless, 
to our knowledge, prior to our work, Dirac electron holography had not been 
reported. Our work provides an initial step in manipulating or controlling Dirac 
waves with the aid of modern machine learning. 

\appendix

\section{Generating data using multiple multipole method}\label{sec:Appendix_A}

\begin{figure}
\centering
\includegraphics[width=\linewidth]{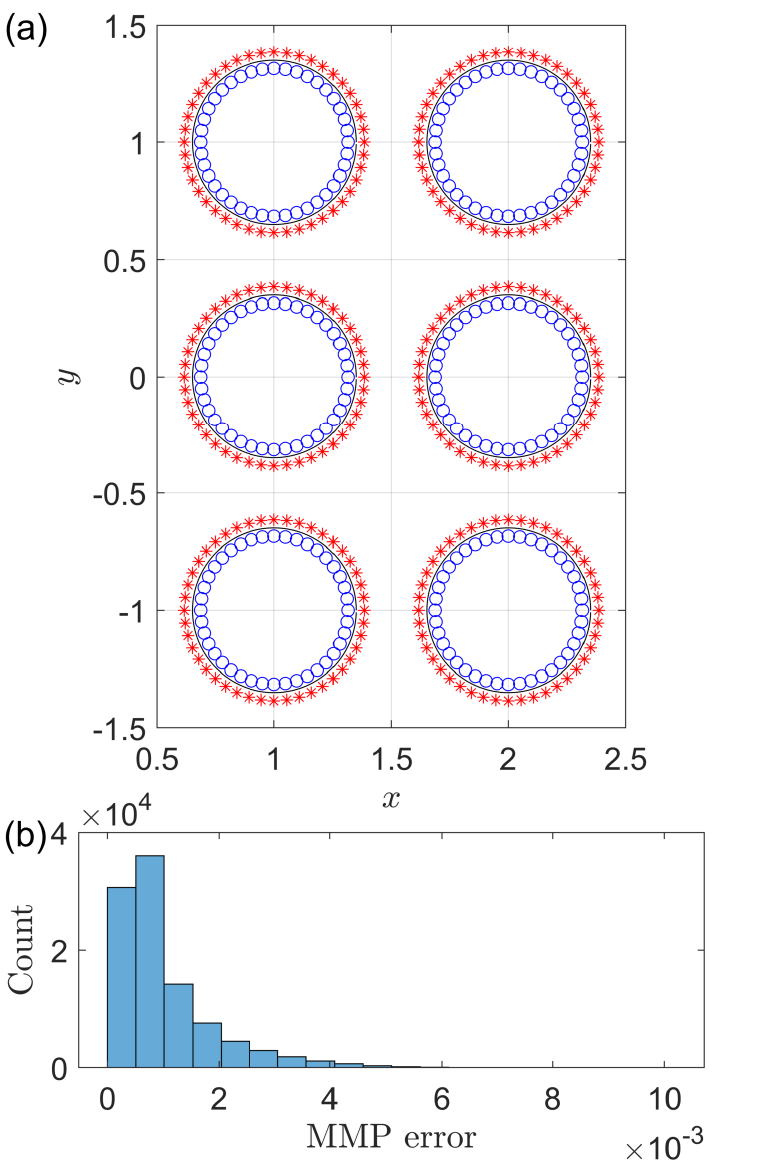}
\caption{MMP method for solving the scattering waveform from a graphene 
metasurface and numerical error. (a) Illustration of the poles inside and outside 
each circular scatter. (b) Histogram of the boundary fitting error. In all cases, 
the values of the energy and potentials are randomly generated.}
\label{fig:MMP}
\end{figure}

Our training data are the scattering waveforms from a sophisticated scatterer 
such as a star-shaped quantum dot, which are generated by the MMP method
originated in optics~\cite{LB:1987,Imhof:1996,KA:2002,MEHV:2002,TE:2004,MEH:2002}
and adopted to Dirac-Weyl spinor systems~\cite{XL:2019,XL:2020,WXL:2020,HXL:2020}.
After a graphene metasurface has been designed, testing it also requires solving
the scattering field from it, which is also done using the MMP method. Take the
metasurface of six quantum dots as an example. The basic idea of MMP is to place 
poles both inside and outside of each boundary, as shown in Fig.~\ref{fig:MMP}(a). 
The wave function outside the metasurface is determined by all the poles inside 
the circular boundaries, and the corresponding poles outside each circular boundary
determine the wave function inside the circle. The detailed computational
procedure can be be found in Ref.~\cite{HXL:2020}. 

\begin{figure}
\centering
\includegraphics[width=\linewidth]{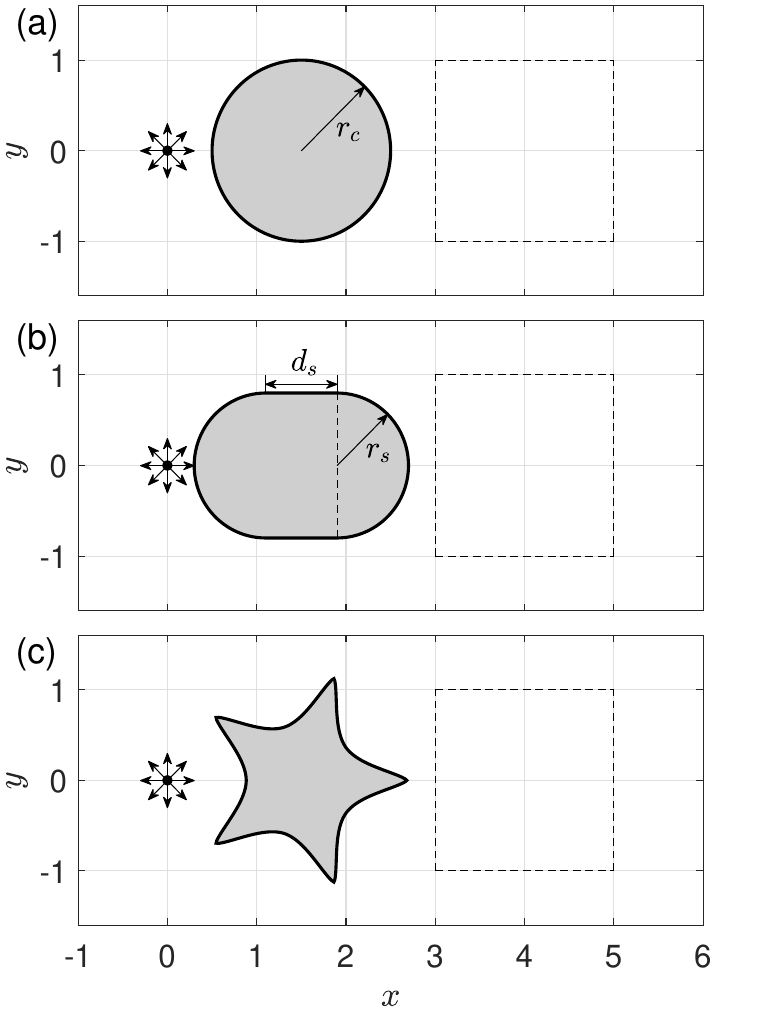}
\caption{Different graphene quantum scatterers for generating desired waveforms. 
(a) A circular quantum dot of radius $r_c=1$. (b) A stadium quantum dot with 
parameters $r_s=l_s=0.8$. (c) A star quantum dot with the boundary determined 
by the Gielis formula.}
\label{fig:geometry_other}
\end{figure}

In our calculation, poles inside each circle are located at $r_i=0.9 r$ at the
same angular interval. The number of poles inside each circle is $N_i=40$. Poles
outside the circle are located at $r_o=1.1 r$ and their number is $N_o=44$. The 
boundary is discretized with $N_j=3(N_i+N_j)$ points. Each pole generates three
values of the angular momentum: $l=\{-1, 0, 1 \}$. For a given metasurface,  
calculating the scattering waveform requires solving roughly $3000$ equations 
with approximately $1500$ unknown parameters. Figure~\ref{fig:MMP}(b) shows 
a typical histogram of the MMP boundary fitting errors~\cite{XL:2019}, which
are sufficiently small to guarantee accurate wave solutions. Altogether, about
$10^5$ waveforms are produced for the results reported in this paper.

\section{Generating target scattering wave} \label{sec:Appendix_B}

The desired target waveforms are calculated using the MMP method for the 
geometrical structures shown in Figs.~\ref{fig:geometry_other}(a-c). The circle 
in Fig.~\ref{fig:geometry_other}(a) has the radius $r_c=1$ and is centered at at 
$[1.5, 0]$. The two geometric parameters defining the stadium scatterer in 
Fig.~\ref{fig:geometry_other}(b) are $r_s=d_s=0.8$ and the center is located 
at $[1.5,0]$. The shape of star scatterer is determined by the Gielis 
formula~\cite{gielis2003generic}:
\begin{align} \label{eq:Gielis}
r(\theta)=\frac{m_3}{\left[\left|\frac{1}{a}\cos\left(\frac{m_1\theta}{4} \right) \right|^{n_1}+\left|\frac{1}{b}\sin\left(\frac{m_2\theta}{4} \right) \right|^{n_2}\right]^{1/n_3}},
\end{align}
with the parameters $m_1=m_2=10$, $m_3=1.2$, $a=0.98$, $b=0.28$, $n_1=2.33$, 
$n_2=1.46$ and $n_3=2.79$. The center of the star is located at $[1.5,0]$.

\section*{Data and code availability}
\vspace*{-0.1in}
Data and codes are available from GitHub: https://github.com/hanchendi/Dirac-Electron-Holography 

\section*{Acknowledgments}
\vspace*{-0.1in}
This work was supported by the Air Force Office of Scientific Research (AFOSR) under Grant No.~FA9550-21-1-0438. VK was supported in part by IDQ (a Quantum Communications Company) and by NSF under Grant No.~NSF-2328991.

\section*{Author Contributions}
\vspace*{-0.1in}
All designed the research project, the models, and methods. C.-D.H.
performed the computations. All analyzed the data. C.-D.H. and Y.-C.L. wrote
the paper. 

\section*{Competing Interests}
\vspace*{-0.1in}
The authors declare no competing interests.

\section*{Correspondence}
\vspace*{-0.1in}
To whom correspondence should be addressed: Ying-Cheng.Lai@asu.edu

\bibliography{graphene_meta}

\end{document}